\documentstyle[12pt]{article}
\begin{document}
\baselineskip=2\baselineskip
\noindent
Running title: {\it Slave-Boson functional-integral
approach to the degenerate Hubbard model}
\begin{center}
{\large\bf Slave-Boson Functional-Integral Approach to}  \\
\end{center}
\begin{center}
{\large\bf  the Hubbard Model with Orbital Degeneracy }  \\
\end{center}

\begin{center}
Hideo Hasegawa$^\dagger$  \\
{\it Department of Physics, Tokyo Gakugei University  \\
Koganei, Tokyo 184, Japan}
\end{center}
\begin{center}
{\rm (Received December 13, 1996)}
\end{center}
\thispagestyle{myheadings}
%
\begin{center} 
{\bf Abstract}   \par
\end{center} 
 
   A slave-boson functional-integral method has been developed for
the Hubbard model with arbitrary, orbital degeneracy $D$.
Its saddle-point mean-field  theory is equivalent 
to the Gutzwiller approximation, 
as in the case of single-band Hubbard model.
Our theory is applied to the doubly degenerate ($D=2$) model,
whose paramagnetic state has been studied by
numerical calculations.
The effect of the exchange interaction on the metal-insulator (MI)
transition is discussed.
The critical interaction for the MI transition is
analytically calculated as  functions of 
orbital degeneracy and  electron occupancy.

\vspace{0.5cm}
\noindent
Keywords: slave boson, Gutzwiller  approximation, metal-insulator 
transition
\vspace{2.0cm}

\noindent
$\dagger$ e-mail address:  hasegawa@u-gakugei.ac.jp
\newpage
\noindent
{\large\bf $\S$1. Introduction}

     The Hubbard model has been accepted as a model for a study
of the strongly correlated systems such as transition metals,
valence-mixing and high-$T_c$ materials.
The single-band Hubbard has been extensively investigated 
by using various methods.$^{1-3)}$
Gutzwiller$^{1)}$ adopted a variational approach, employing the
variational wave function.  Since the exact evaluation
of the ground-state energy is difficult, he used an additional
approximation which is now called the Gutzwiller approximation (GA).
A validity of the GA was investigated by Monte-Carlo technique
for finite-size clusters.$^{4,5)}$
It is realized that the GA becomes a  better approximation for the 
higher-dimensional system.$^{6,7)}$  
The GA has been widely employed for various studies
including magnetism and the metal-insulator (MI) transition.$^{8)}$

  Kotliar and Ruckenstein (KR)$^{9)}$ proposed the slave-boson 
functional-integral method, whose saddle-point approximation
is shown to be equivalent to the GA.
The KR method has a wider applicability than the GA 
as follows:

\noindent
(1) within the saddle-point theory, we can deal with 
the complicated magnetic systems such as antiferromagnetic
state besides paramagnetic and ferromagnetic states.$^{10,11)}$

\noindent
(2) we can go beyond the saddle-point approximation, by 
including fluctuations around the saddle points.$^{12-16)}$

\noindent
(3) we can extend the theory to finite temperatures.$^{11,17)}$

  In contrast with the single-band Hubbard model, little studies
based on the GA had been reported 
on the  degenerate-band Hubbard model 
since its first application was made
by Gutzwiller and Chao.$^{18)}$
In the last few years, however,
several attempts$^{19-21)}$ were reported of re-formulating
and applying the GA to degenerate-band Hubbard model.
Lu$^{19)}$ obtained the analytical expression of 
the critical interaction for the MI transition. 
Okabe$^{20}$ proposed a sophisticated method 
in calculating the band-narrowing factor, 
{\it q}, which is the most difficult  part in applying the GA.
Quite recently B\"{u}nemann and Weber$^{21)}$ discussed the 
first-order MI transition in the half-filled doubly degenerate
Hubbard model when the exchange interaction is included.

  It is  desirable to develop the slave-boson functional-integral 
theory for the degenerate-band Hubbard model,
which is the purpose of the present paper.
We have developed such a  theory, adopting the slave-boson method
of Dorin and Schlottman$^{22)}$ which
was originally employed for the Anderson lattice model.
As in the single-band Hubbard model, the saddle-point approximation
to our slave-boson theory is equivalent to the GA 
in the degenerate-band  Hubbard model, and our slave-boson theory
has the above-mentioned advantages 
with the wider applicability than the GA.

  The paper is organized as follows:  In the next $\S$2,
we  present a slave-boson functional-integral formulation for the
Hubbard model with the $D$-fold orbital degeneracy.  
Our formalism is applied to the doubly degenerate $(D = 2)$ model, 
for which numerical calculations 
are performed and reported in $\S$3.
Section 4 is devoted to conclusion and supplementary discussion. 
In the Appendix we analytically calculate the critical 
interaction strength for the MI transition, 
in order to demonstrate the feasibility of our theory.
\vspace{1.0cm}
\noindent
{\large\bf $\S$2. Formulation}

\noindent
{\bf $\S$2.1 Basic Equations}
    
   We adopt the Hubbard model with $D$-fold orbital degeneracy whose
Hamiltonian is given by
\begin{equation}
H = \sum_{\sigma} \sum_{i j} \sum_{m m'} t^{m m'}_{ij} 
c^\dagger_{im \sigma} c_{jm' \sigma}
+ \frac{1}{2} \sum_i \sum_{(m,\sigma) \neq (m',\sigma')}
U_{mm'}^{\sigma\sigma'}
c^\dagger_{im \sigma} c_{im \sigma}
c^\dagger_{im' \sigma'} c_{im' \sigma'},
\end{equation}
\noindent
where $c_{im\sigma}$ is an annihilation operator of an electron 
with an orbital index {\it m} 
and spin $\sigma (= \uparrow, \downarrow)$ on the lattice site $i$
and $t_{ij}^{mm'}$ is the hopping integrals.
As for the on-site interaction, $U_{mm'}^{\sigma \sigma'}$, 
we introduce three parameters:
$U_0$  for $m = m', \sigma \neq \sigma'$, 
$U_1$ for $m \neq m', \sigma \neq \sigma'$, and
$U_2$ for $m \neq'm', \sigma = \sigma'$. 
We assume that the spin and orbital degrees of freedom are not
coupled for simplicity.

   We employ the slave boson formulation of Dorin and Schlottman,$^{22)}$
introducing $2^{2D}$ bose annihilation operators:
\begin{equation}
b^{(\ell)}_{i;m_1 \sigma_1,m_2 \sigma_2, ...., m_{\ell}\sigma_{\ell}},
\end{equation}
which project to the configuration of $\ell$ electrons with pairs of
orbital and spin indices $\{m \sigma \}$. In eq. (2) all the index pairs are 
different because of the Pauli principles.
  
We introduce the following product operations:$^{22)}$

\noindent
(1) A full contraction is defined as
\begin{equation}
\left( {\sl b}_i^{(\ell) \dagger} \cdot {\sl b}_i^{(\ell)} \right) 
\equiv \sum_{ m_1 \sigma_1,m_2 \sigma_2, ...., m_{\ell}\sigma_{\ell}}
b^{(\ell) \dagger}_{i;m_1 \sigma_1,m_2 \sigma_2, ...., m_{\ell}\sigma_{\ell}} \:
b^{(\ell)}_{i;m_1 \sigma_1,m_2 \sigma_2, ...., m_{\ell}\sigma_{\ell}}.
\end{equation}
\noindent
(2) A partial contraction with a fixed pair of $(m_n\sigma_n)$ 
for $\ell \geq 1$ is defined as
\begin{equation}
({\sl b}_i^{(\ell) \dagger} \cdot {\sl b}_i^{(\ell)})_{m_n\sigma_n} 
\equiv \sum_{ m_1 \sigma_1,m_2 \sigma_2, ...., m_{n-1}\sigma_{n-1},
m_{n+1} \sigma_{m+1}1, ...., m_{\ell}\sigma_{\ell}}
b^{(\ell) \dagger}_{i;m_1 \sigma_1,m_2 \sigma_2, ...., m_{\ell}\sigma_{\ell}} \:
b^{(\ell)}_{i;m_1 \sigma_1,m_2 \sigma_2, ...., m_{\ell}\sigma_{\ell}}.
\end{equation}
\noindent
With the use of these operations the completeness of the 
slave boson operators is given by 
\begin{equation}
\sum^{2D}_{\ell = 0}
({\sl b}_i^{(\ell) \dagger} \cdot {\sl b}_i^{(\ell)}) = 1,
\end{equation}
while the correspondence between bosons and fermions is expressed by
\begin{equation}
c^\dagger_{im \sigma} c_{im' \sigma} 
 = \sum^{2D}_{\ell = 1}
({\sl b}_i^{(\ell) \dagger} \cdot {\sl b}_i^{(\ell)})_{m \sigma}
= n_{im\sigma}.
\end{equation}

  The partition function of the model given by eq. (1) is expressed 
as a functional integral over coherent states of Fermi and Bose fields.
The constraints given by eqs. (5) and (6) are incorporated with
Lagrange multipliers, $\lambda^{(1)}$ and $\lambda_\sigma^{(2)}$, to get 
\begin{equation}
Z = \int \Pi_{\ell=0}^{2D} Db^{(\ell)} \int D \lambda^{(1)} 
\int D\lambda^{(2)}_\sigma \:\:
exp [- \int^{\beta}_0 d\tau \:\: (L_f(\tau) + L_b(\tau))],
\end{equation}
where the actions, $L_f$ and $L_b$, are given by
\begin{equation}
L_f(\tau) = \sum_{i} \sum_{m\sigma} \left[ c^\dagger_{im\sigma} 
\left( \partial / \partial \tau + \lambda_{im\sigma}^{(2)} \right)
c_{im\sigma} \right]
+ \sum_{\sigma} \sum_{i j} \sum_{m m'} t^{m m'}_{ij} 
z^\dagger_{im\sigma} c^\dagger_{im \sigma} c_{jm' \sigma} z_{jm\sigma}, 
\end{equation}
%
%
%
\begin{eqnarray}
L_b(\tau) =  
\sum_i \left[ \sum^{2D}_{\ell=0} b_i^{\ell \dagger} 
\left( \partial/ \partial \tau 
+ \lambda_i^{(1)}\right) b_i^{\ell} 
-  \sum_{m\sigma} \sum_{\ell = 1}^{2D} \lambda^{(2)}_{im\sigma}
(b_i^{(\ell) \dagger} \cdot b_i^{(\ell)})_{m\sigma} 
- \lambda_i^{(1)} \right] + \Phi_0,
\end{eqnarray}
\begin{equation}
\Phi_0 = \sum_i \; \sum_{\ell=2}^{2D} \;
\sum_{ m_1 \sigma_1,m_2 \sigma_2, ...., m_{\ell}\sigma_{\ell}} \;\;
\sum_{(m\sigma,m'\sigma')} \;\; U_{mm'}^{\sigma\sigma'} \;\;
b^{(\ell) \dagger}_{i;m_1 \sigma_1,m_2\sigma_2, ...., m_{\ell}\sigma_{\ell}} \:
b^{(\ell)}_{i;m_1 \sigma_1,m_2\sigma_2, ...., m_{\ell}\sigma_{\ell}},
\end{equation}
the fourth summation in eq. (10) being made over a pair of 
indices $(m\sigma,m'\sigma')$ 
with $(m\sigma) \neq (m'\sigma')$
in the configuration: 
$\{m_1\sigma_1,m_2\sigma_2,...,m_{\ell}\sigma_{\ell}\}$
occupied by $\ell$ electrons.
It is noted that when $U_0 = U_1 = U_2 = U$, we get
$\Phi_0 =  U \; \sum_i \:\sum^{2D}_{\ell=2} C^\ell_2 \; 
({\sl b}_i^{(\ell) \dagger} \cdot  {\sl b}_i^{(\ell) })$
where $C_k^{\ell} = \ell!/k! \: (\ell-k)!$.  
Although a choice of the factor, $z_{im\sigma}$, is not unique, 
the expression given by 
\begin{equation}
z_{im\sigma } = (1 - n_{im\sigma})^{-1/2}
\sum^{2D}_{\ell=1}  
\left( {\sl b}_i^{(\ell-1) \dagger} \cdot  {\sl b}_i^{(\ell)} \right)_{m\sigma}
n_{im\sigma}^{-1/2},
\end{equation}
is shown to yield the correct result in the limit of vanishing 
interactions.$^{22)}$

We employ the static approximation for boson fields.
Furthermore we adopt the  following change of variables:
\begin{equation}
\nu_{im } = (\lambda^{(2)}_{im\uparrow}
+ \lambda^{(2)}_{im\downarrow})/2, \:\:
\xi_{im } = - (\lambda^{(2)}_{im\uparrow}
- \lambda^{(2)}_{im\downarrow})/2,
\end{equation}
\begin{equation}
n_{im} = <n_{im\uparrow}> + <n_{im\uparrow}>, \:\: 
m_{im} = <n_{im\uparrow}> - <n_{im\uparrow}>,
\end{equation}
where the brackets, $< \:\: >$,  stand for the average.
Thus the expression of the partition function becomes
\begin{equation}
Z = \int \Pi_{\ell=2}^{2D} Db^{(\ell)} 
\int D\nu \int D\xi \int Dn \int Dm 
\:\: exp \: (- \beta  \Phi),
\end{equation}
where
\begin{equation}
e^{- \beta \Phi} = exp \left( - \beta
\left[ \sum_i 
\sum_m (\xi_{im} m_{im}-\nu_{im} n_{im}) 
+ \Phi_0 \; \right] \right)  \:
Tr \:\: exp \: (- \beta H_{eff}). 
\end{equation}
The effective one-electron
Hamiltonian, $H_{eff}$,  is given by 
\begin{equation}
H_{eff} = \sum_{\sigma} \sum_{i j} \sum_{m m'} 
z_{im\sigma} z_{jm'\sigma}
t^{m m'}_{ij} \: 
c^\dagger_{im \sigma} c_{jm' \sigma}
+ \sum_\sigma \sum_i \sum_m 
(\nu_{im} - \sigma \xi_{im}) \:
c^\dagger_{im \sigma} c_{im \sigma},
\end{equation}
%
%
%
\begin{equation}
z_{im\sigma } = 2 (2 - n_{im})^{-1/2}
n_{im}^{-1/2}
\left[ \surd \overline{e_i \; p_{im\sigma}}  
+ \surd \overline{p_{im\sigma}} \;\; b_i^{(2)} 
+ \sum^{2D}_{\ell=3}  
\left( {\sl b}_i^{(\ell-1) } \cdot  
{\sl b}_i^{(\ell)} \right)_{m\sigma} \right],
\end{equation}
where $e_i$ and $p_{im\sigma}$ are given by
\begin{equation}
e_i = ({\sl b}_i^{(0)} \cdot {\sl b}_i^{(0)})
= 1 - \sum_{m} n_{im} + 
\sum_{m\sigma}
\sum_{\ell=2}^{2D} \: \:
[(\ell - 1)/\ell] \:\:
({\sl b}_i^{(\ell)} \cdot {\sl b}_i^{(\ell)})_{m\sigma},
\end{equation}
\begin{equation}
p_{im\sigma} = ({\sl b}_i^{(1)} \cdot {\sl b}_i^{(1)})_{m\sigma}
= (n_{im} + \sigma m_{im})/2 - \sum_{\ell=2}^{2D} 
({\sl b}_i^{(\ell)} \cdot {\sl b}_i^{(\ell)})_{m\sigma}.
\end{equation}
It should be noted that the functional integral given by eqs. (14)-(19)
is performed over variables of 
$\xi, \nu, n, m,$ and $b^{(\ell)}$ with $\ell \ge 2$.
Our  expression has  physically more transparent meaning than 
an original  static approximation in which
the functional integral is carried out over the variables of
$\lambda^{(1)}, \lambda_\sigma^{(2)}$ and 
$b^{(\ell)}$ with $\ell \ge 0$, 
although both the approaches are equivalent.
The expression for the functional integral given by eqs. (14)-(19)
is a simple generalization of the single-band model to 
the degenerated-band model.$^{11)}$
When all the integration variables are replaced by their
saddle-point values, we get the mean-field approximation,
which is expected to be equivalent with the GA theory,$^{19-21)}$ 
related discussions being given in $\S$4.

\vspace{1.0cm}
\noindent
{\bf $\S$2.2 Doubly degenerate band}

  We apply our theory developed so far to the doubly degenerate  Hubbard model.
After a simple calculation,  the expressions for 
the functional integral given by 
eqs. (14)-(19) become
\begin{equation}
Z = \int Dd_{0}\int Dd_{1}\int Dd_{\sigma}\int Dt_\sigma\int 
Df\int D\xi\int Dm\int D\nu\int Dn 
\:\: exp [- \beta ( \Phi_0+\Phi_1+\Phi_2)],
\end{equation}
\begin{equation}
\Phi_0 = \sum_i \left[ 2 U_0 d_{i0} + 2 U_1 d_{i1} 
+ U_2 (d_{i\uparrow} + d_{i\downarrow})
+ 2(U_0 + U_1 + U_2 ) (t_{i\uparrow} + t_{i\downarrow}+ f_i) \right],
\end{equation}
\begin{equation}
\Phi_1 = \sum_{im} \left[ \xi_{im} m_{im} + (\mu - \nu_{im}) n_{im} \right],
\end{equation}
\begin{equation}
\Phi_2 = \int d\varepsilon f(\varepsilon) (- 1/\pi) Im \:\:
Tr \:\: ln \:  G(\varepsilon).
\end{equation}
where $f(\varepsilon)$ is the Fermi distribution function
and $\mu$ Fermi level.
The one-particle Green function, $G(\varepsilon)$, is expressed in terms of
the effective Hamitonian, $H_{eff}$, for the system under
the charge ($\nu_{im}$) and exchange ($\xi_{im}$) fields, as
\begin{equation}
G(\varepsilon) = (\varepsilon - H_{eff})^{-1},
\end{equation}
with
\begin{equation}
H_{eff} = \sum_{\sigma} \sum_{i j} \sum_{m m'} 
z_{im\sigma} 
t^{m m'}_{ij} 
z_{jm'\sigma} \;
c^\dagger_{im \sigma} c_{jm' \sigma}
+ \sum_\sigma \sum_i \sum_m 
(\nu_{im} - \sigma \xi_{im}) \:
c^\dagger_{im \sigma} c_{im \sigma},
\end{equation}
%
\begin{eqnarray}
z_{im\sigma} =  \frac{ 2 \left[ 
\surd \overline{p_{i\sigma}} 
( \surd \overline{e_{i}} + \surd \overline{d_{i\sigma}} )
+ ( \surd \overline{d_{i0}} + \surd \overline{d_{i1}} )
( \surd \overline{p_{i-\sigma}} + \surd \overline{t_{i\sigma}} )
+ \surd \overline{t_{i-\sigma}} 
( \surd \overline{d_{i-\sigma}} + \surd \overline{f_{i}} ) \right] }
{(n_{im}+\sigma m_{im})^{1/2} \:
(2 - n_{im} - \sigma m_{im})^{1/2}  },
\end{eqnarray}
\begin{equation}
e_i = 1 - 2 n_{im} + 2d_{i0} + 2 d_{i1} 
+ d_{i\uparrow} + d_{i\downarrow}
+ 4 (t_{i\uparrow} + t_{i\downarrow}) + 3 f_i,
\end{equation}
\begin{equation}
p_{i\sigma} = (n_{im} + \sigma m_{im})/2  
- (d_{i0} + d_{i1} + d_{i\sigma}) 
- (2 t_{i\sigma} + t_{i-\sigma}) -  f_i,
\end{equation}
where $t_{i\sigma}$ and $f_i$ stand for  
the $\ell = 3$ and 4 components of  
$({\sl b}_i^{(\ell)} \cdot {\sl b}_i^{(\ell)})_{m\sigma} $, 
respectively.  As for $\ell = 2$, we take into account 
the three kinds of configurations:
$d_{i0}$ for a pair of electrons on the same orbital with opposite spin, 
$d_{i1}$ on the different  orbital with opposite spin,  and
$d_{i\sigma}$ on the different  orbital with same spin $\sigma$. 

The explicit form of the Green function, $G(\varepsilon)$, 
depends on the electronic and magnetic structures
of a system to be investigated.
Since the effective transfer integrals in eq. (25) is expressed as
a product form: $z_{im\sigma} t_{ij}^{mm'} z_{jm'\sigma}$,
we can express the one-electron Green function in terms of 
the locators defined by$^{23)}$
\begin{equation}
X_{im\sigma} = (\varepsilon - \nu_{im} + \sigma \xi_{im})
/r_{im\sigma},
\end{equation}
where $r_{im\sigma}=(z_{im\sigma})^2$, 
with which the band-narrowing factor,
$q_{ij\sigma}^{mm'}$, is expressed as
\begin{equation}
q_{ij\sigma}^{mm'} \equiv z_{im\sigma} z_{jm'\sigma}
=\surd \overline{r_{im\sigma} \; r_{jm'\sigma}},
\end{equation}

The mean-field free energy can be obtained from the saddle-point values of the
integration variables for which 
the variational conditions yield the following simultaneous equations:
\begin{equation} 
n_{im} = \sum_\sigma <c^\dagger_{im\sigma} c_{im\sigma}>,
\end{equation}
\begin{equation} 
m_{im} = \sum_\sigma \sigma <c^\dagger_{im\sigma} c_{im\sigma}>,
\end{equation}
\begin{equation} 
\mu - \nu_{im} + \sum_\sigma R_{im\sigma} 
(\partial r_{im\sigma} /\partial n_{im} ) = 0,
\end{equation}
\begin{equation} 
\xi_{im} + \sum_\sigma R_{im\sigma} 
(\partial r_{im\sigma} /\partial m_{im} ) = 0,
\end{equation}
\begin{equation}
2 U_0 + \sum_{m \sigma} R_{im\sigma} 
(\partial r_{im\sigma}/\partial d_{i0})  = 0,
\end{equation}
\begin{equation}
2 U_1 + \sum_{m \sigma} R_{im\sigma} 
(\partial r_{im\sigma}/\partial d_{i1})  = 0,
\end{equation}
\begin{equation}
U_2 +  \sum_{m \sigma'} R_{im\sigma'} 
(\partial r_{im\sigma'}/\partial d_{i\sigma})  = 0,
\end{equation}
\begin{equation}
2(U_0 + U_1 + U_2) +  \sum_{m \sigma'} R_{im\sigma'} 
(\partial r_{im\sigma'}/\partial t_{i\sigma})  = 0,
\end{equation}
\begin{equation}
2 (U_0 + U_1 + U_2) + \sum_{m \sigma} R_{im\sigma} 
(\partial r_{im\sigma}/\partial f_i)  = 0,
\end{equation}
where $R_{im\sigma} = \partial \Phi_2/\partial r_{im\sigma}$.

  In the remainder of this paper we consider 
only the paramagnetic states at $T = 0$ K, for which we get
\begin{equation}
\Phi_2 = \int^\mu d\varepsilon \:\: (\varepsilon - \mu) \:\:
(- 1/\pi) \: Im \:\: Tr \: G(\varepsilon),
\end{equation}
\begin{equation}
Tr \: G(\varepsilon) = 
\sum_\sigma \sum_m \frac{1}{\left( \varepsilon - \nu 
-  q \varepsilon_{k} \right) },
\end{equation}
\begin{equation}
\varepsilon_0(n) =  \: \sum_\sigma R_\sigma, 
\end{equation}
\begin{equation}
q = r = \frac{4  \left[ ( \surd \overline{d_{0}}
+  \surd \overline{d_{1}} + \surd \overline{d_{2}} )
(\surd \overline{p} + \surd \overline{t} ) + \surd \overline{e p} 
+ \surd \overline{t f}) \right]^2 }
{n (2 - n)},
\end{equation}
with
\begin{equation}
e = 1 - 2 n + 2(d_{0} + d_{1} + d_{2}) + 8 t + 3 f,
\end{equation}
\begin{equation}
p = n/2  - (d_{0} + d_{1} + d_{2}) - 3 t -  f.
\end{equation}
Here $d_2 = d_\sigma$, $t = t_\sigma$ 
and $R_{\sigma} = R_{im\sigma}$ {\it et al.},
subscripts $i$ and $m$ being omitted;
$\varepsilon_{k}$ is the Fourier transform of the transfer integrals
and $\varepsilon_0(n)$ is the  ground-state energy 
per sub-band for $U_0 = U_1 = U_2 = 0$  as a function of
$n$, the number of electrons per sub-band.

The ground-state energy per site is then given by
\begin{equation}
E = 2 q \varepsilon_0(n) 
+ 2 U_0 d_{0} + 2 U_1d_{1} + 2 U_{2} d_{2}
+ 2 (U_0 + U_1 + U_2) (2t + f), 
\end{equation}
which is identical with that obtained by Bl\"{u}menann and 
Weber.$^{21)}$ Employing the  \\ 
Gutzwiller-type wave function,
they obtained the ground-state energy given by eq. (46),
from which the optimum values of the occupancies are determined
by the variational conditions for $E$ given by
\begin{equation}
U_0 + \varepsilon_0  (\partial q/\partial d_{0})  = 0,
\end{equation}
\begin{equation}
U_1 + \varepsilon_0  (\partial q/\partial d_{1})  = 0,
\end{equation}
\begin{equation}
U_2 + \varepsilon_0 (\partial q/\partial d_{2})  = 0,
\end{equation}
\begin{equation}
2(U_0 + U_1 + U_2) +  \varepsilon_0 (\partial q/\partial t)  = 0,
\end{equation}
\begin{equation}
(U_0 + U_1 + U_2) + \varepsilon_0 (\partial q/\partial f)  = 0.
\end{equation}
It is easy to see that eqs. (35)-(39) reduce to eqs. (47)-(51)
in the paramagnetic states.

\vspace{1.0cm}
\noindent
{\large\bf $\S$3. Numerical Calculations}

     Numerical calculations have been performed for $D=2$ model
with the quarter-filled 
($n = 0.5$) and half-filled ($n = 1$) bands, which are considered to
be most interesting.

     We firstly show the calculated results by taking 
$U_0 = U_1 = U_2 = U$, for which $d_0, d_1$ and $d_2$ are equivalent:
$d_0 = d_1 = d_{2} =  d$.
The $U$-dependence of the occupancies for $n = 0.5$ is shown in Fig. 1.
At $U = 0$ we get $e = 0.3164, p= 0.1055, d = 0.0352,
t = 0.0117$ and $f = 0.0039$.
When $U$ value is increased, $e, \: t$ and $f$ decrease while $p$ increases.
In particular, $t$ and $f$ become at least ten times smaller
than $e, \: p$ or $d$ at $U/\mid \! \varepsilon_0(n) \! \mid > 3$.
The q factor also decreases and vanishes above $U_c$ which
stands for the critical interaction for the metal-insulator (MI) 
transition.  Near the MI transition, $q$ behaves as 
$q \propto (U_c - U)$ as in the single-band model.$^{8)}$
We get $U_c = 13.2 \mid \! \varepsilon_0(n) \! \mid$, 
which agrees with the previous results.$^{19,21)}$
Above $U_c$ only $p$ remains finite with a value of 0.25.
The $U$ dependence of the ground-state energy is shown in Fig. 2,
where the energy calculated by the Hartree-Fock  approximation (HFA):
\begin{equation}
E_{\rm HF} = 2 \varepsilon_0(n) + (1/2)(U_0 + U_1 + U_2)\:\:n^2,
\end{equation}
is also plotted for a comparison. 
The  ground-state energy in the GA is always 
lower than $E_{\rm HF}$ as expected.

    Figure 3 shows the $U$ dependence of the occupancies 
for half-filled ($n = 1.0$) case, in which the relation:
\begin{equation}
e = f, \:\:  p = t,
\end{equation}
holds by the electron-hole symmetry. 
When $U$ value is increased, 
only $d$ increases while other occupancies decrease.
The band-narrowing factor, $q$, vanishes above $U_c$, which is
$12.0 \mid \varepsilon_0(n) \mid$  in agreement with the result of
refs. 19 and 21.
The double occupancy $d$ persists even above $U_c$ with a value of 0.167,
which is in contrast with the quarter-filled case where
only single occupancy $p$ survives. 
We might expect that in the limit of infinite $U$, any states
with more than doubly occupied state vanish. This seems not the case for the
half-filled $D = 2$ degenerate band, where doubly occupied state 
may survive by making a compromise with the constraints given by the
completeness of bosons (eq. (5))
and the electron-hole symmetry (eq. (53)).
This situation is changed when the exchange interaction is introduced,
as will be discussed shortly.
Fig. 4 shows the ground-state energy of the half-filled case
as a function of $U$.
Again the energy obtained in our theory is lower than that
in the HFA.

     Next we take into account the exchange interaction, $J$,
with which our three interaction parameters are expressed as
\begin{equation}
U_0 = U, \:\:\:\: U_1 = U' - J = U - 2J, \:\:\:\: 
U_2 = U' - J = U - 3 J,
\end{equation}
where we employ the relation: $U - U' = 2J$  
required for the rotational symmetry,$^{24)}$ 
$U$ and $U'$ being the intra- and inter-band Coulomb interactions,
respectively.

Figures 5 and 6 show the $U$ dependence of the occupancies 
and the ground-state energy, respectively,
for the quarter-filled band with $J/U = 0.1$.
The general behavior shown in Figs. 5 and 6 is similar to
that in Figs. 1 and 2, respectively,
except the fact that the degeneracy among doubly occupied states
is removed with an inclusion of the exchange interaction, $J$.
Because $U_0 > U_1 > U_2$ for finite $J$, we get
$d_0 < d_1 < d_{2}$.
We realize that an inclusion of the exchange interaction also enhances
the value of $U_c$:
$U_c/ \mid \varepsilon(n) \mid$ = 13.2, 16.4, and 24.8
for $J/U$ = 0.0, 0.1 and 0.2, respectively. 
The MI transition is of the second order in all the  
quarter-filled cases investigated, 

     On the contrary, the effect of exchange interaction 
is quite different in the case of half-filled band.
The MI transition becomes the {\it first-order} transition as is shown
in Fig. 7, where the $U$ dependence of the occupancies with $J/U = 0.1$
for $n= 1.0$ is plotted.
This first-order MI transition has been recently pointed out 
by B\"{u}nemann an Weber.$^{21)}$
A change from the second-order to first-order MI transition
is induced with an infinitesimally small inclusion of $J$.
When $J = 0$, all doubly occupies states are equivalent and
remain finite even at $U > U_c$, as was shown in Fig. 3.
For a finite $J$, however, $d_{2}$ is the most favorable
state among the three doubly occupied states, and
only $d_{2}$ survives
above $U = 6.0$, where neither $d_0$ nor $d_1 $ remain finite.
This first-order transition is realized in the $U$ dependence of
the ground-state energy shown in Fig. 8, where the $E - U$ curve of
the metallic state does not tangentially
intersect that of the insulating state at the MI transition,
in contrast with the case of  the second-order transition shown 
in Fig. 2, 4 and 6.

\vspace{1.0cm}
\noindent
{\large\bf $\S$4. Conclusion and Discussion}

     To summarize, we have developed the slave-boson  
functional-integral
theory for the Hubbard model with the orbital degeneracy $D$, 
employing
the slave-boson method of Dorin and Schllotman$^{22)}$ 
who  originally applied it to the Anderson model.  
Our slave-boson mean-field theory has been used 
for a study of the doubly
degenerate model, for which numerical calculations have been performed.
We have shown that our slave-boson mean-field theory is equivalent with
the GA for $D = 2$.$^{21)}$ 
It is shown in the Appendix that our mean-field theory and Lu's$^{19)}$ 
yield the identical critical interaction
of the MI transition for arbitrary $D$.
This suggests that the equivalence between the slave-boson mean-field
theory and the GA$^{19-21)}$ generally holds for any degeneracy $D$,
although it has not been  rigorously proved yet.

     We expect that our slave-boson mean-field theory is a convenient 
and useful method in studying strongly correlated systems described 
by the degenerate-band Hubbard model.
With the use of the Green function method, we can deal with the 
complicated magnetic  states within the mean-field approximation.
It would be indispensable to investigate the antiferromagnetic states$^{25)}$
for a deeper understanding of the MI transition of the $D=2$ Hubbard
model.
Our slave-boson mean-field theory enables us to investigate 
the occupation versus interaction phase diagram 
of the degenerate Hubbard model,
which was previously studied within the Hartree-Fock approximation.
Furthermore, we can take account of fluctuations around the saddle points,
going beyond the mean-field approximation.
It is interesting to extend our theory such as to include the
effect of spin fluctuations at finite temperatures
in order to discuss the finite-temperature magnetism of transition metals.

\vspace{1.0cm}
\noindent{\bf Acknowledgment}  \par
\vspace{0.2cm}

This work is partly supported by
a Grant-in-Aid for Scientific Research from the Japanese 
Ministry of Education, Science and Culture.

\newpage
\vspace{1.0cm}
\noindent
{\large\bf $\S$ Appendix}

   As one of applications of our formalism, we calculate the critical
interaction, $U_c$,  for the MI transition of the Hubbard with the orbital
degeneracy, $D$, in the paramagnetic states.
We assume that near the second-order 
MI transition point, fluctuations of the electron 
occupancy around its  average $N$ (integer) are small.
Then we take into
only three possible states: $N - 1, N$ and $N +1$ states, whose occupancies 
are expressed by $e, p$ and  $d$, respectively, after Lu.$^{19)}$
 
  The constraints given by eqs. (5) and (6) become
\begin{eqnarray}
C^{2D}_{N-1} \:e + C^{2D}_N \: p + C^{2D}_{N+1} \: d = 1,    
\;\;\;\;\;\;\;\;\;\;\;\;\;\;\;\;\;\;\;\;\;\;\; 
\mbox{(A$\cdot$1)} \nonumber  
\end{eqnarray}
\begin{eqnarray}
C^{2D-1}_{N-2} \: e + C^{2D-1}_{N-1} \: 
p + C^{2D-1}_{N} \: d = n/2,              
\;\;\;\;\;\;\;\;\;\;\;\;\;\;\;\;\;\;\;\;\;\;\; 
\;\;\;\;\;\;\;\;\; \mbox{(A$\cdot$2)} \nonumber
\end{eqnarray}
where $n \:\; (= N/D$) is the number of electrons per sub-band and
$C_k^\ell \:\; [= \ell !/k! \: (\ell-k)!$ ] is zero when
either $\ell $, $k$, or $(\ell-k)$ is negative as convention.
The ground-state energy when $U_0 = U_1 = U_2 = U $ in eq. (1)
is given by
\begin{eqnarray}
E = q \: D \: \varepsilon_0(n) + U \left[ C^{2D}_{N-1} C^{N-1}_2 \: e  
+ C^{2D}_{N} C^N_2 \: p + C^{2D}_{N+1} C^{N+1}_2 \: d
\right],                                                    
\;\;\;\;\;\;\;\;\; 
\;\;\;\;\;\;\;\;\; \mbox{(A$\cdot$3)} \nonumber
\end{eqnarray}
where $q, e$ and $p$ are given as a function of $d$ as 
\begin{eqnarray}
q = [4/n(2-n)] \: [C^{2D-1}_{N-1} \surd \overline{e \: p }
+ C^{2D-1}_N \surd \overline{p \: d} ]^2.                 
\;\;\;\;\;\;\;\;\;\;\;\;\;\;\;\;\;\;\;\;\;\;\; 
\;\;\;\;\;\;\;\;\; \mbox{(A$\cdot$4)} \nonumber
\end{eqnarray}
\begin{eqnarray}
e = ( \Delta_1/\Delta) \: d,                             
\;\;\;\;\;\;\;\;\;\;\;\;\;\;\;\;\;\;\;\;\;\;\; 
\;\;\;\;\;\;\;\;\; \mbox{(A$\cdot$5)} \nonumber
\end{eqnarray}
\begin{eqnarray}
p = (\Delta_2/\Delta) - (\Delta_3/\Delta) \: d,         
\;\;\;\;\;\;\;\;\;\;\;\;\;\;\;\;\;\;\;\;\;\;\; 
\;\;\;\;\;\;\;\;\; \mbox{(A$\cdot$6)} \nonumber
\end{eqnarray}
with
\begin{eqnarray}
\Delta = C^{2D-1}_{N-1} C^{2D}_{N-1} - C^{2D-1}_{N-2} C^{2D}_{N},
\;\;\;\;\;\;\;\;\;\;\;\;\;\;\;\;\;\;\;\;\;\;\; 
\;\;\;\;\;\;\;\;\; \mbox{(A$\cdot$7)} \nonumber
\end{eqnarray}
\begin{eqnarray}
\Delta_1 = C^{2D-1}_{N} C^{2D}_{N} - C^{2D-1}_{N-1} C^{2D}_{N+1},
\;\;\;\;\;\;\;\;\;\;\;\;\;\;\;\;\;\;\;\;\;\;\; 
\;\;\;\;\;\;\;\;\; \mbox{(A$\cdot$8)} \nonumber
\end{eqnarray}
\begin{eqnarray}
\Delta_2 = (n/2) \: C^{2D}_{N-1} - C^{2D-1}_{N-2},              
\;\;\;\;\;\;\;\;\;\;\;\;\;\;\;\;\;\;\;\;\;\;\; 
\;\;\;\;\;\;\;\;\; \mbox{(A$\cdot$9)} \nonumber
\end{eqnarray}
\begin{eqnarray}
\Delta_3 = C^{2D-1}_{N} C^{2D}_{N-1} - C^{2D-1}_{N-2} C^{2D}_{N+1}.
\;\;\;\;\;\;\;\;\;\;\;\;\;\;\;\;\;\;\;\;\;\;\; 
\;\;\;\;\;\;\;\;\; \mbox{(A$\cdot$10)} \nonumber
\end{eqnarray}
Taking the variational condition for  $E$ with respect to $d$, we obtain
the critical interaction, $U_c$, given by
\begin{eqnarray}
\frac{U_c}{\mid\varepsilon_0(n)\mid}
= \frac{D}{N(2D-N)}
\left[\surd \overline{N (2D - N + 1)} 
+ \surd \overline{(2 D -N)(N + 1)}
\right]^2,                                                     
\;\;\;\;\;\;\;\;\; \mbox{(A$\cdot$11)} \nonumber
\end{eqnarray}
which is identical with the result obtained by Lu.$^{19)}$
Equation (A$\cdot$11) is symmetric with respect to the interchange of
the variable: $N \leftrightarrow (2D - N)$ as it should be.
In the limit of infinite $D$ and finite $N$, eq. (A$\cdot$11) becomes
\begin{eqnarray}
\lim_{D \rightarrow \infty}
\left[ \frac{U_c}{D \mid\varepsilon_0(n)\mid} \right]
= \frac{1}{N}
\left[ \surd \overline{N} + \surd \overline{N + 1}
\right]^2.                                                     
\;\;\;\;\;\;\;\;\; \mbox{(A$\cdot$12)} \nonumber
\end{eqnarray}
Equation (A$\cdot$12) is in agreement with the result of ref. 26, in which
the fermion problem is treated as the boson problem using the GA.
Some calculations using Eq. (A$\cdot$11) are shown in Fig. 9.

\newpage
\newpage

\newpage
\noindent{\large\bf  Figure Captions}   \par
\vspace{1.0cm}
\noindent
{\bf Fig. 1} 
The occupancies, $e, p, d, t$ and $f$, and the
band narrowing factor, $q$, as a function of $U$ for the 
quarter-filled ($n = 0.5$) case with $J =0$, 
the results of $t $ and
$f$ being multiplied by a factor of ten. 
\vspace{0.5cm}

\noindent
{\bf Fig. 2} 
The ground-state energy for  $n = 0.5$ and $J = 0$
in the GA (solid curve) and the 
HFA (dashed curve),
the arrow denoting the MI transition.
\vspace{0.5cm}

\noindent
{\bf Fig. 3} 
The occupancies, $e, p, d, t$ and $f$, and the
band narrowing factor, $q$, as a function of $U$ for the 
half-filled ($n = 1.0$) case with $J = 0$.
\vspace{0.5cm}

\noindent
{\bf Fig. 4} 
The ground-state energy for  $n = 1.0$ and $J = 0$
in the GA  (solid curve) and the 
HFA (dashed curve),
the arrow denoting the MI transition.
\vspace{0.5cm}

\noindent
{\bf Fig. 5} 
The occupancies and the
band narrowing factor, $q$, as a function of $U$ for the 
quarter-filled ($n = 1/2$) case with $J/U = 0.1$,
the results of $t $ and
$f$ being multiplied by a factor of ten. 
\vspace{0.5cm}

\noindent
{\bf Fig. 6} 
The ground-state energy for  $n = 0.5$ and $J/U = 0.1$
in the GA  (solid curve) and the 
HFA (dashed curve),
the arrow denoting the MI transition.
\vspace{0.5cm}

\noindent
{\bf Fig. 7} 
The occupancies and the
band narrowing factor, $q$, as a function of $U$ for the 
half-filled ($n = 1.0$) case with $J/U = 0.1$.
\vspace{0.5cm}

\newpage
\noindent
{\bf Fig. 8} 
The ground-state energy for  $n = 1.0$ with $J/U = 0.1$
in the GA  (solid curve) and the 
HFA (dashed curve),
the dotted curve showing the energy of the insulating state ($q = 0$)
and the arrow denoting the MI transition.
\vspace{0.5cm}

\noindent
{\bf Fig. 9} 
The critical interaction for the MI transition, 
$U_c/\mid \! \varepsilon_0(n) \! \mid$,
against the electron occupancy per sub-band ($n = N/D$)  of the
Hubbard model with $N $ electrons and $D$-fold orbital degeneracy.
Only points shown by filled circles are meaningful,
dashed curves being drawn using eq. (A$\cdot$11).

\end{document}